\documentclass[article,aps,showpacs,nofootinbib,prd]{revtex4-1}
\usepackage{amssymb}
\usepackage{amsmath}
\usepackage[utf8]{inputenc}
\usepackage{graphics}
\usepackage{epsfig}
\usepackage{enumerate}

\begin{document}

\title{Casimir Energy in a Bounded Gross-Neveu model}

\author{F. Escalante}
\email{fescalante@ucn.cl}
\affiliation{Departamento de F\'{\i}sica, Universidad Cat\'olica del Norte, Angamos 0610, Antofagasta, Chile}

\author{J.C. Rojas}
\email{jurojas@ucn.cl}
\affiliation{Departamento de F\'{\i}sica, Universidad Cat\'olica del Norte, Angamos 0610, Antofagasta, Chile}

\begin{abstract}
In this letter we study some relevant physical parameters of the massless  Gross-Neveu (GN) model in a finite spatial dimension for different boundary conditions.
It is considered the standard homogeneous Hartree Fock solution
using zeta function regularization for the study the mass dynamically generated and its respective beta function. It is found that the beta function does not depend on the boundary conditions.
On the other hand, it was considered the Casimir effect of the resulting effective theory. There appears a complex picture where the sign  of the generated forces depends on the parameters used in the study.

\end{abstract}

\maketitle



\section{Introduction}

The Gross-Neveu (GN) model was born as a toy model of Quantum Chromodynamics (QCD) \cite{Gross:1974jv}. Despite its simplicity, it keeps many interesting features, such as asymptotic freedom, dynamical mass generation and discrete  chiral symmetry.
Later, it was used in the study of baryons with explicit symmetry breaking by a mass term \cite{Thies:2005vq}.

Curiously, this model has also application in condensed matter physics, where it describes the conductivity in
certain polymers. In particular, it can be mentioned the case of
{\cal trans}-polyacetylene, which, in a simplified continuous model, is described by the symmetric GN model \cite{saxena}, besides, the massive GN model has a condensed matter analogue; which are polymers with non-degenerate ground states \cite{brazovskii}.

The original treatment of the GN model was under the assumption
of the unbroken translational invariance,
it means an standard treatment based on the large $N$ approximation,
 where the use of the Hartree Fock (HF) approximation is well founded,
 that leads a
condensate independent of the space coordinates. Later,
it was realized that there are crystal solutions of the model
i.e. an spatial realization solution
which have a rich interpretation in the realm of condensed
matter physics \cite{Thies:2005wv}.


In our study, we shall concentrate on the homogeneous solutions of
the GN model for a finite space of fixed size $L$. We are interested in the behaviour of physical parameters for different boundary conditions (BC's). 

The spatial BC's considered are the periodic , anti periodic conditions.
There are also considered the situation of no current transmission 
on the borders, there we consider two cases where such condition is fulfilled (see appendix \ref{apC}).

The HF approximation, implies the use of a large momentum cutoff.
Since we shall deal with systems of spacial finite size, the momentum integrals must be replaced by summation on discrete modes, meaning that the natural regularization to be used is the zeta regularization technique \cite{Kirsten:2010zp}.

In this work, we first ask about the ultraviolet dependence of the physical parameters on the BC's, considering the GN model 
at zero bare mass ($m_0=0$) where temperature and chemical potential are not considered. 
We assume that the spatial length $L$ is a fixed parameter, so,
if the physical mass is independent of the cutoff, it implies that the beta function does not depend on the BC's. There appears an arbitrary mass scale and the functional dependency of the dynamical mass clearly depends on the BC's.

A second step in our work is to study the Casimir energy and force due to the quantum fluctuation of the effective free system that arises from the HF approximation. We consider the non dimensional parameter 
$\mu =mL$, since the value of m is fixed by ultraviolet considerations, the variation of $\mu$ is equivalent to the variation of $L$. We find that the value of energy and Force are sensitive to the BC's. In particular, the signature of the energy clearly differs in the small size limit, but it is universally negative for infinite size limit.
On the other hand, the force is also sensitive to the BC's, implying situations where the forces are such that they compress or expand our space depending on the BC's used. There is also a universal metastable point where the force becomes zero independently of the BC's used. For the large $L$ limit the force becomes positive for any BC's considered.


\section*{The Gross-Neveu model}\label{GN}

The Gross-Neveu Lagrangian is given by

\begin{equation}
\label{lagrangian}
\mathcal{L}_{GN}=\bar{\psi}^{i}i\gamma^{\mu}\partial_{\mu}\psi^{i}+\frac{1}{2}g^{2}\left\lbrace
(\bar{\psi}^{i}\psi^{i})^{2}-\lambda (\bar{\psi}^{i}\gamma_5\psi^{i})^{2}
\right\rbrace
-m_0\bar{\psi}^{i}\psi^{i}.
\end{equation}

\noindent Where $i$ runs from 1 to $N$, it was introduce a finite mass in order to consider a general expression and we use the convention

\[
\gamma^0=\left( \begin{matrix}
0 & -1 \\
-1 & 0
\end{matrix}  \right), \; \gamma^1=\left( \begin{matrix}
i & 0 \\
0 & -i
\end{matrix}  \right), \; \gamma^5=\left( \begin{matrix}
0 & i \\
-i & 0
\end{matrix}  \right). \;
\]

The Euler Lagrange equation from (\ref{lagrangian}) is given by

\begin{equation}
\label{EL}
i\gamma^{\mu}\partial_{\mu}\psi+g^{2}\left\lbrace
\bar{\psi}\psi^{i}-\lambda (\bar{\psi}^{j}\gamma_5\psi^{j}) \gamma_5
\right\rbrace \psi^{i}-m_0\psi^{i}=0.
\end{equation}

For the sake of simplicity, from now, we suppress the index $i$. In the framework of Hartree-Fock relativistic approximation, it is assumed the expectation value $\langle
\bar{\psi}\gamma_5\psi\rangle=0$ and $\langle
\bar{\psi}\psi\rangle=N\rho$. We end up with the expression

\begin{equation}
\label{dirac}
(i\gamma^{\mu}\partial_{\mu}-m)\psi (x)=0,
\end{equation}

\noindent where $m=m_0-g^2 N\rho$ and $\rho=\langle \bar{\psi}\psi  \rangle/N$.

\noindent From (\ref{dirac}) we obtain a free Dirac equation

\[
i\frac{\partial \psi}{\partial t}=H\psi=\begin{pmatrix} 0 & -i \\ i & 0 \end{pmatrix} \left(-i\partial_x\right) \psi
-m \begin{pmatrix} 0 & 1 \\ 1 & 0 \end{pmatrix} \psi.
\]

\noindent In order to obtain a stationary solution, we use the usual decomposition

\[
\psi(x) = {\rm e}^{-i\lambda t}\displaystyle{\phi(x) \choose \chi(x)}.
\]

\noindent We obtain

\begin{equation}
\lambda  {\phi \choose \chi}=-i\begin{pmatrix} 0 & i \\ -i & 0 \end{pmatrix} {\partial_{x}\phi  \choose  \partial_{x}\chi} + \begin{pmatrix} 0 & -m \\ -m & 0 \end{pmatrix}{\phi  \choose \chi},
\end{equation}

\noindent giving a system of coupled equations

\begin{eqnarray}
\frac{d}{dx}\chi(x)-m\chi(x)& = &\lambda\phi(x),\\
-\frac{d}{dx}\phi(x)-m\phi(x)& = &\lambda\chi(x).
\end{eqnarray}

\noindent By making the redefinition of the fields

\begin{equation}
\label{fg}
f  =  \chi+\phi,\;
g  =  \chi-\phi,
\end{equation}

\noindent we obtain a general solution

\begin{eqnarray}
\label{sgf}
f(x)& = &\frac{\alpha}{\sqrt{\lambda+m}}\cos(\Omega x) - \frac{\beta}{\sqrt{\lambda+m}}\sin(\Omega x),\\
\label{sgg}
g(x)& = &\frac{\alpha}{\sqrt{\lambda-m}}\sin(\Omega x) + \frac{\beta}{\sqrt{\lambda-m}}\cos(\Omega x),
\end{eqnarray}

\noindent where $\Omega=\sqrt{\lambda^{2}-m^{2}}$ ande the constants $\alpha$ and $\beta$ are not independent since they are determined by
the boundary conditions.


\section{Hartree Fock for different boundary conditions}\label{BC}

Following the standard procedure \cite{Schon:2000he}, it is possible to compute the negative energy in an infinite space taking the value of $m$ as a parameter to be determined

\begin{equation}
\label{den}
\frac{\mathcal{E}}{N}=-2\int_{\mid k \mid \leq \Lambda}\frac{dk}{2\pi}\sqrt{m^{2}+k_{n}^{2}}+\frac{m^{2}}{2G},
\end{equation}

\noindent where $G=Ng^{2}$ and $\Lambda$ is a momentum cut off.

Since we have a finite spatial size, the wave number $k$ is discretized $k_n=(2\pi n+\phi)/r L$, implying   $\int dk/(2\pi) \to \frac{1}{r L}\sum$, Where $r$ is a number which depends on boundary conditions. So, we have

\begin{equation}
\label{disc}
\frac{\mathcal{E}}{N}=-\frac{2}{r L}\sum_{n} \left(m^{2}+k_{n}^{2}\right)^{1/2} +\frac{m^{2}}{2G}.
\end{equation}

\noindent 
The summation term can be expressed as generalized zeta function regularization and its result is described in appendix A. Since the power $1/2$ in the summation is replaced by a term $1/2-\epsilon$, it appears a mass scale $\eta$. We have the following momentum decomposition for the BC to be considered (section (\ref{spc})):

\[
\begin{matrix} {\rm Periodic} & k_n=2\pi n/L & n \in (-\infty,\infty), \\ 
{\rm Antiperiodic} & k_n=(2n+1)\pi /L & n \in (-\infty,\infty), \\
{\rm Zero\:current\; i)} & k_n=(2n)\pi/2L & n \in (-\infty,\infty),\\
{\rm Zero\:current\; ii)} & k_n=(2n+1)\pi /2L & n \in (-\infty,\infty).
 \end{matrix} 
\]

\noindent For the four considered BC, we obtained the following expressions for the energy density,  where it was introduced the non dimensional variables $\mu=mL$ and $\tilde{\eta}=\eta L$ (see appendix \ref{apB}):

\begin{itemize}

\item Periodic BC

\begin{equation}
\label{energia.hf.p}
\frac{\mathcal{E}^P}{N}=-\frac{\mu^{2}}{2\pi\epsilon L^{2}}+\frac{\mu^{2}}{2\pi L^{2}}-\frac{\mu^{2}}{\pi L^{2}}\ln\left(\frac{2\tilde{\eta}}{\mu}\right)+\frac{4\mu}{\pi L^{2}}\sum_{n=1}^{\infty}\frac{K_{1}(\mu n)}{n}+\frac{\mu^{2}}{2G L^{2}}.
\end{equation}

\item Anti periodic BC

\begin{equation}
\label{energia.hf.ap}
\frac{\mathcal{E}^{AP}}{N}=-\frac{\mu^{2}}{2\pi\epsilon L^{2}}+\frac{\mu^{2}}{2\pi L^{2}}-\frac{\mu^{2}}{\pi L^{2}}\ln\left(\frac{2\tilde{\eta}}{\mu}\right)+\frac{4\mu}{\pi L^{2}}\left\{\sum_{n=1}^{\infty}\frac{K_{1}(2\mu n)}{n}-\sum_{n=1}^{\infty}\frac{K_{1}(\mu n)}{n}\right\}+\frac{\mu^{2}}{2G L^{2}}.
\end{equation}

\item Zero current  BC

\begin{eqnarray}
\label{energia.hf.zci}
\frac{\mathcal{E}^{i}}{N} &=& -\frac{\mu^{2}}{2\pi\epsilon L^{2}}+\frac{\mu^{2}}{2\pi L^{2}}-\frac{\mu^{2}}{\pi L^{2}}\ln\left(\frac{2\tilde{\eta}}{\mu}\right)+\frac{2\mu}{\pi L^{2}}\sum_{n=1}^{\infty}\frac{K_{1}(2\mu n)}{n}+\frac{\mu^{2}}{2G L^{2}}, \\
\frac{\mathcal{E}^{ii}}{N} &=& -\frac{\mu^{2}}{2\pi\epsilon L^{2}}+\frac{\mu^{2}}{2\pi L^{2}}-\frac{\mu^{2}}{\pi L^{2}}\ln\left(\frac{2\tilde{\eta}}{\mu}\right)+\frac{2\mu}{\pi L^{2}}\sum_{n=1}^{\infty}(-1)^{n}\frac{ K_{1}(2\mu n)}{n}+\frac{\mu^{2}}{2GL^{2}}.
\end{eqnarray}

\end{itemize}

\noindent In the following step, we minimize the energy densities with respect to $\mu$. Then, we use (\ref{G}) and obtain
for each BC an expression for $G$ 

\begin{eqnarray}
\label{gp}
G^P &=& \pi\left\{ \frac{1}{\epsilon}-2+2\ln\left(\frac{2\tilde{\eta}}{\mu}\right)+4\sum_{n=1}^{\infty}K_{0}(n\mu)\right\}^{-1},
\\
\label{gap}
G^{AP} &=& \pi\left\{\frac{1}{\epsilon}-2+2\ln\left(\frac{2\tilde{\eta}}{\mu}\right)-4\sum_{n=1}^{\infty}(K_{0}(n\mu)-2K_{0}(2n\mu))\right\}^{-1}, \\
\label{gi}
G^{i} &=& \pi\left\{\frac{1}{\epsilon}-2+2\ln\left(\frac{2\tilde{\eta}}{\mu}\right)+4\sum_{n=1}^{\infty}K_{0}(2n\mu)\right\}^{-1}, \\
\label{gii}
G^{ii} &=& \pi\left\{\frac{1}{\epsilon}-2+2\ln\left(\frac{2\tilde{\eta}}{\mu}\right)+4\sum_{n=1}^{\infty}(-1)^{n}K_{0}(2\mu n)\right\}^{-1}, 
\end{eqnarray}

\noindent where $\epsilon=s+1/2$ goes to zero and must be considered as the ultraviolet cut-off.

If we Consider $\mu$ and $\tilde{\eta}$ constants, so the running
of G should depends on the BC's.
But, 
fixing the value of a common $G$ for certain scale, implying different values of $\mu$ for each BC. If we take the limit $\epsilon\rightarrow 0$, we observe an universal behaviour for G, independent of the BC's

We observe from the general relation (\ref{G}), that there is  dependency of the
constants $\mu$ and $\tilde{\eta}$ for each BC through the transcendental equation:

\[
2\ln\left(\frac{2\tilde{\eta}}{\mu}\right)+\frac{4}{\pi}\sum_{n=1}^{\infty}\cos(\phi n)K_{0}(\mu r n)=C,
\]

\noindent being $C$ an arbitrary constant.
 
Considering the traditional point of view where the physical $\mu$ must be independent of the cutt-off
$\epsilon$ we have a renormalization group equation 

\[
\epsilon^2 \frac{d\mu}{d \epsilon}=
\epsilon^2 \frac{\partial \mu}{\partial  \epsilon}+
\epsilon^2 \frac{\partial  G}{\partial  \epsilon} \frac{\partial \mu}{\partial  G}+
\epsilon^2 \frac{\partial  \tilde{\eta}}{\partial  \epsilon} \frac{\partial \mu}{\partial  \tilde{\eta}}
=0.
\]

\begin{figure}[h]
\centering
\includegraphics[width=60mm,scale=0.5]{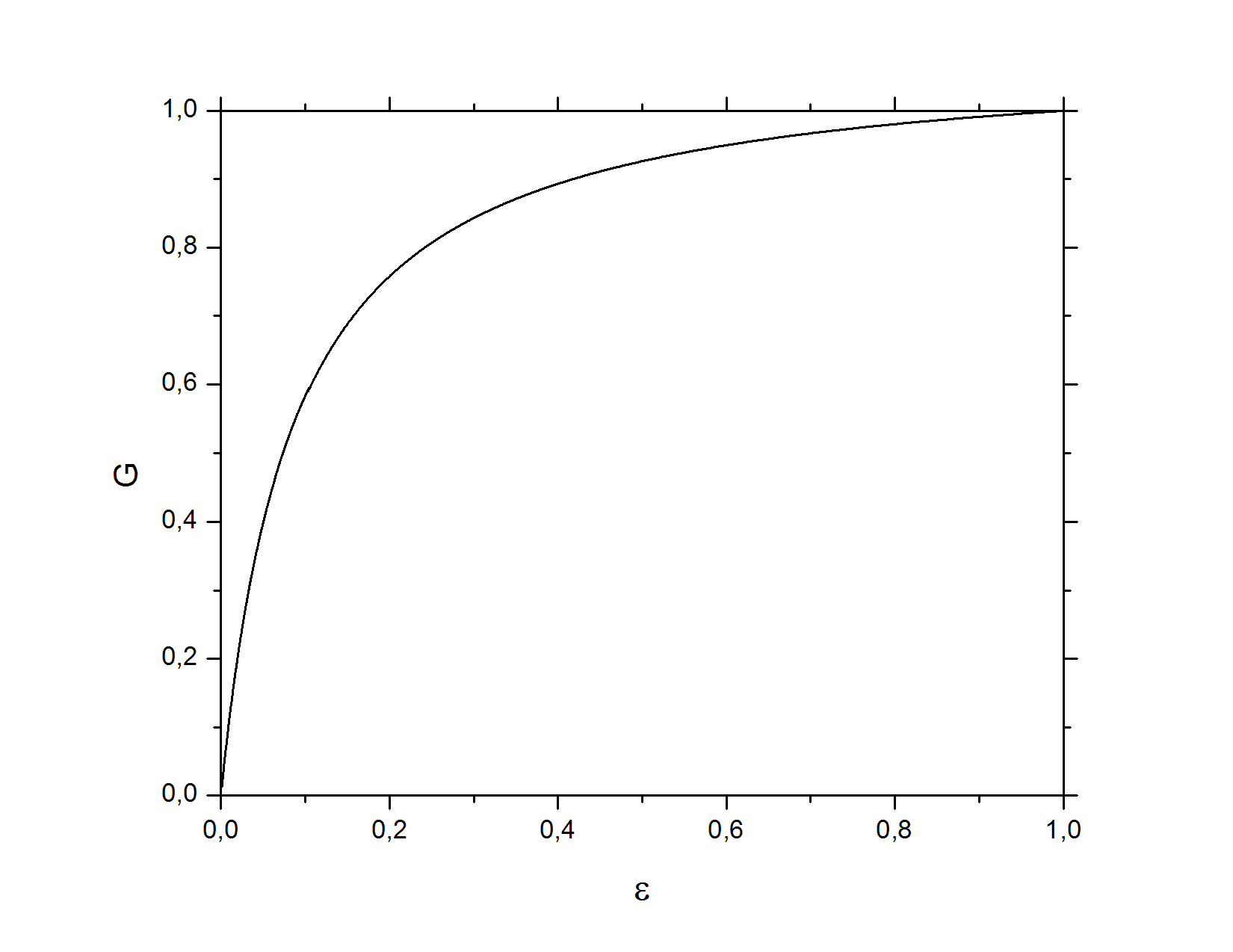}
\caption{The running of G, for different BC's fixing the parameters in order to have $G=1$ for $\epsilon=1$.}
\label{GG}
\end{figure}

\noindent For (\ref{gap})-(\ref{gii}), it is computed the beta function

\[
\beta=\epsilon^2\frac{d G}{d \epsilon},
\]

\noindent we  obtain

\begin{equation}
\label{beta.ap}
\beta^{P} =\beta^{AP}=\beta^{i} = \beta^{ii}= \frac{G^{2}}{\pi}, \\
\end{equation}

\noindent meaning an universal behaviour of $G(\epsilon)$ as it is shown if figure (\ref{GG}).

\section{Casimir Energy for global boundary conditions}

Imposing BC's of the form

\[
{\phi(x+L) \choose \chi(x+L)} = {\rm e}^{i\alpha}{\phi(x) \choose \chi(x)}.
\]

\noindent It is equivalent to study

\[
{f(x+L) \choose g(x+L)} = {\rm e}^{i\alpha}{f(x) \choose g(x)},
\]

\noindent because of the linear relation (\ref{fg}).

In equations (\ref{sgf}) and (\ref{sgg}) we have solutions of the form

\begin{equation}
 {f(x) \choose g(x)} =  \begin{pmatrix} f_{1}(x) & f_{2}(x) \\ g_{1}(x) & g_{2}(x) \end{pmatrix} {\alpha \choose \beta},
\end{equation}

\noindent where the values of $\alpha, \beta$ depend on the imposed BC on the problem. We can define the matrix

\begin{equation}
\mathbb{H}(x)=\begin{pmatrix} f_{1}(x) & f_{2}(x) \\ g_{1}(x) & g_{2}(x) \end{pmatrix}.
\end{equation}

\noindent Assuming that it is invertible, i.e. $\det(\mathbb{H}(x))\ne 0$, we can isolate the constants

\begin{equation}
 {\alpha \choose \beta} = \mathbb{H}^{-1}(0) {f(0) \choose g(0)},
\end{equation}

\noindent meaning that

\begin{equation}
 {f(x) \choose g(x)}=\mathbb{H}(x)\mathbb{H}^{-1}(0) {f(0) \choose g(0)}.
 \label{gev}
\end{equation}

\noindent On the other side, the BC can be expressed in the following way

\begin{equation}
 {f(L) \choose g(L)} =  \mathbb{M} {f(0) \choose g(0)},
 \label{gBC}
\end{equation}

\noindent so, evaluating (\ref{gev}) in $x=L$ and comparing with
(\ref{gBC}), we have

\begin{equation}
 \left[  \mathbb{M}- \mathbb{H}(L)\mathbb{H}^{-1}(0)  \right]{f(0) \choose g(0)}
 \Rightarrow  \det \left[  \mathbb{M}- \mathbb{H}(L)\mathbb{H}^{-1}(0)  \right]=0.
  \label{sgav}
\end{equation}

\noindent
Which is the condition for the eigenvalues of the problem.

We can include the periodic and anti periodic case by the parametrization

\[
\mathbb{M}=\begin{pmatrix} \exp\left(i\alpha \right) & 0  \\ 0 & \exp\left(i\alpha \right) \end{pmatrix}.
 \]

\noindent From (\ref{sgf}) and (\ref{sgg}), we have

\[
\mathbb{H}(x)=
\begin{pmatrix}  \frac{\cos(\Omega x)}{\sqrt{\lambda+m}}
& -\frac{\sin(\Omega x)}{\sqrt{\lambda+m}}  \\
\frac{\sin(\Omega x)}{\sqrt{\lambda-m}} &
\frac{\cos(\Omega x)}{\sqrt{\lambda-m}} \end{pmatrix}.
 \]

\noindent Eq. (\ref{sgav}) leads to the condition

\[
\cos \Omega L =\cos \phi \rightarrow \Omega^2=\frac{(2\pi n+\phi)^2}{r^{2}L^2},\;n \in \mathbb{Z}.
\]

\noindent Since $\Omega=\sqrt{\lambda^2-m^2}$, we have

\begin{equation}
\lambda_n^2=m^2+\frac{(2\pi n+\phi)^2}{r^{2}L^2}
= \frac{4\pi^2}{r^{2}L^2}\left[\frac{m^2L^2r^{2}}{4\pi^2}+ \left(n+\frac{\phi}{2\pi}\right)^2\right].
\label{conditiong}
\end{equation}

\noindent We define $\mu=mL$ and $r$  a parameter which depends of the boundary conditions, so, the general expression for the Casimir energy is given by

\[
E_{Cas}=\langle \hat{H} \rangle=\frac{1}{2}\sum_n \lambda_n
=\lim_{s\rightarrow -1/2} \frac{1}{2}\left(\frac{2\pi}{rL}\right)^{-2s}\sum_{n=-\infty}^{\infty}
\left[\frac{\mu^2r^{2}}{4\pi^2}+ \left(n+\frac{\phi}{2\pi}\right)^2\right]^{-s}. 
\]

From appendix A, we have

\begin{eqnarray}
E_{Cas} &=& \frac{1}{2}{\rm FP}\left(\frac{4\pi^2}{r^{2}L^2}\right)^{1/2-\epsilon}\left(\frac{1}{\eta^2}\right)^{-\epsilon}
\left[
\frac{\mu^2 r^{2}}{8\pi^2\epsilon}-\frac{\mu^ 2 r^{2}}{8\pi^2}
-\frac{\mu^ 2 r^{2}}{4\pi^2}\ln\frac{\mu r}{4\pi}-\frac{\mu}{\pi^2}
\sum_{n=1}^{\infty}\frac{\cos\phi n}{n}
K_1(\mu r n)
\right], \noindent \\
&=& \frac{\pi}{r L}{\rm FP} \left(\frac{\eta L r}{2\pi}\right)^{2\epsilon} \left[
\frac{\mu^2 r^{2}}{8\pi^2\epsilon}-\frac{\mu^ 2 r^{2}}{8\pi^2}
-\frac{\mu^ 2 r^{2}}{4\pi^2}\ln\frac{\mu r}{4\pi}-\frac{\mu}{\pi^2}
\sum_{n=1}^{\infty}\frac{\cos\phi n}{n}
K_1(\mu r n)
\right], \\
&=&  \frac{\pi}{r L} {\rm FP}\left[
\frac{\mu^2 r^{2}}{8\pi^2\epsilon}-\frac{\mu^ 2 r^{2}}{8\pi^2}
+\frac{\mu^ 2 r^{2}}{4\pi^2}\ln\frac{2\eta L}{\mu}-\frac{\mu r}{\pi^2}
\sum_{n=1}^{\infty}\frac{\cos\phi n}{n}
K_1(\mu r n)
\right].
\end{eqnarray}

\noindent Ending with

\begin{equation}
\label{ecas}
{\xi}_{Cas}\equiv LE_{Cas}=
-\frac{\mu^ 2 r}{8\pi}
+\frac{\mu^ 2 r}{4\pi}\ln\frac{2\eta L}{\mu }-\frac{\mu}{\pi}
\sum_{n=1}^{\infty}\frac{\cos\phi n}{n}
K_1(\mu r n).
\end{equation}

The Casimir Force

\begin{eqnarray}
F_{Cas} &=& -\frac{d  E_{Cas}}{d L}=\frac{{\xi}_{Cas}}{L^2}
-\frac{1}{L}
\frac{\partial {\xi}_{Cas}}{\partial L}- \frac{m}{L}\frac{\partial {\xi}_{Cas}}{\partial \mu}, \nonumber \\
&=& -\frac{\mu^2 r}{4\pi L^2}+\frac{{\xi}_{Cas}}{L^2}
- \frac{\mu}{L^2}\frac{\partial {\xi}_{Cas}}{\partial \mu}, \nonumber \\
&=& \frac{\mu^2 r}{8\pi L^2}
-\frac{\mu^ 2 r}{4\pi L^2}\ln \frac{2\eta L}{\mu}
-\frac{\mu^2 r}{\pi L^2}
\sum_{n=1}^{\infty} \cos(\phi n)
K_0(\mu r n)
-\frac{\mu}{\pi L^2} \sum_{n=1}^{\infty} 
\frac{\cos(\phi n)}{n}K_1(\mu r n).
\end{eqnarray}

We define $\tilde{\eta}\equiv  \eta L$, and ${\cal F}\equiv F/m^{2}$, so

\begin{equation}
\mathcal{F}_{Cas}\equiv\frac{F_{Cas}}{m^{2}}=
\frac{r}{8\pi}
-\frac{r}{4\pi}\ln \frac{2\tilde{\eta}}{\mu}
-\frac{r}{\pi}
\sum_{n=1}^{\infty} \cos(\phi n)
K_0(\mu r n)
-\frac{1}{\mu\pi}
\sum_{n=1}^{\infty} \frac{\cos(\phi n)}{n}
K_1(\mu r n).
\end{equation}


\section{Specific boundary conditions}
\label{spc}

\noindent\textbf{Anti Periodic BC}

\bigskip

\bigskip

We first, assume anti periodic BC for our spinor solution

\[
\label{con.ap}
\begin{pmatrix}f(x+L)  \\ g(x+L) \end{pmatrix}
=-\begin{pmatrix}f(x)  \\ g(x) \end{pmatrix} =
\begin{pmatrix} -1 & 0  \\ 0 & -1 \end{pmatrix}
\begin{pmatrix} f(x) \\ g(x) \end{pmatrix}.
\]

\noindent So, we have $\phi=\pi$ in (\ref{conditiong}), so

\begin{equation}
\label{auto.ap}
\lambda_{n}^{2}=m^{2}+ \frac{(2\pi n+\pi)^{2}}{L^{2}} 
\rightarrow r=1.
\end{equation}

\noindent Taking non-dimensional parameter $\mu=mL$, using (\ref{ecas}) we have the Casimir energy

\begin{equation}
\label{energia.ap}
{\xi}_{Cas}^{AP}=-\frac{\mu^{2}}{8\pi}
+\frac{\mu^{2}}{4\pi}\ln\left(\frac{2\tilde{\eta}}{\mu}\right)
+\frac{\mu}{\pi}
\sum_{n=1}^{\infty}\left(\frac{K_{1}\mu n)}{n}-\frac{K_{1}(2\mu n)}{n}\right),
\end{equation}
 
\noindent and the Casimir force

\begin{equation}
\label{fuerzaap}
\mathcal{F}^{AP}_{Cas}=\frac{1}{8\pi}
-\frac{1}{4\pi}\ln\left(\frac{2\tilde{\eta}}{\mu}\right)
+\frac{1}{\pi}\sum_{n=1}^{\infty}\left(K_{0}(\mu n)-2K_{0}(2\mu n)\right)
+\frac{1}{\mu\pi}\sum_{n=1}^{\infty}\left(\frac{K_{1}(\mu n)}{n}-\frac{K_{1}(2\mu n)}{n}\right).
\end{equation}


\begin{figure}[h]
\hfill
\begin{minipage}[t]{.45\textwidth}
\begin{center}
\includegraphics[width=100mm,scale=0.5]{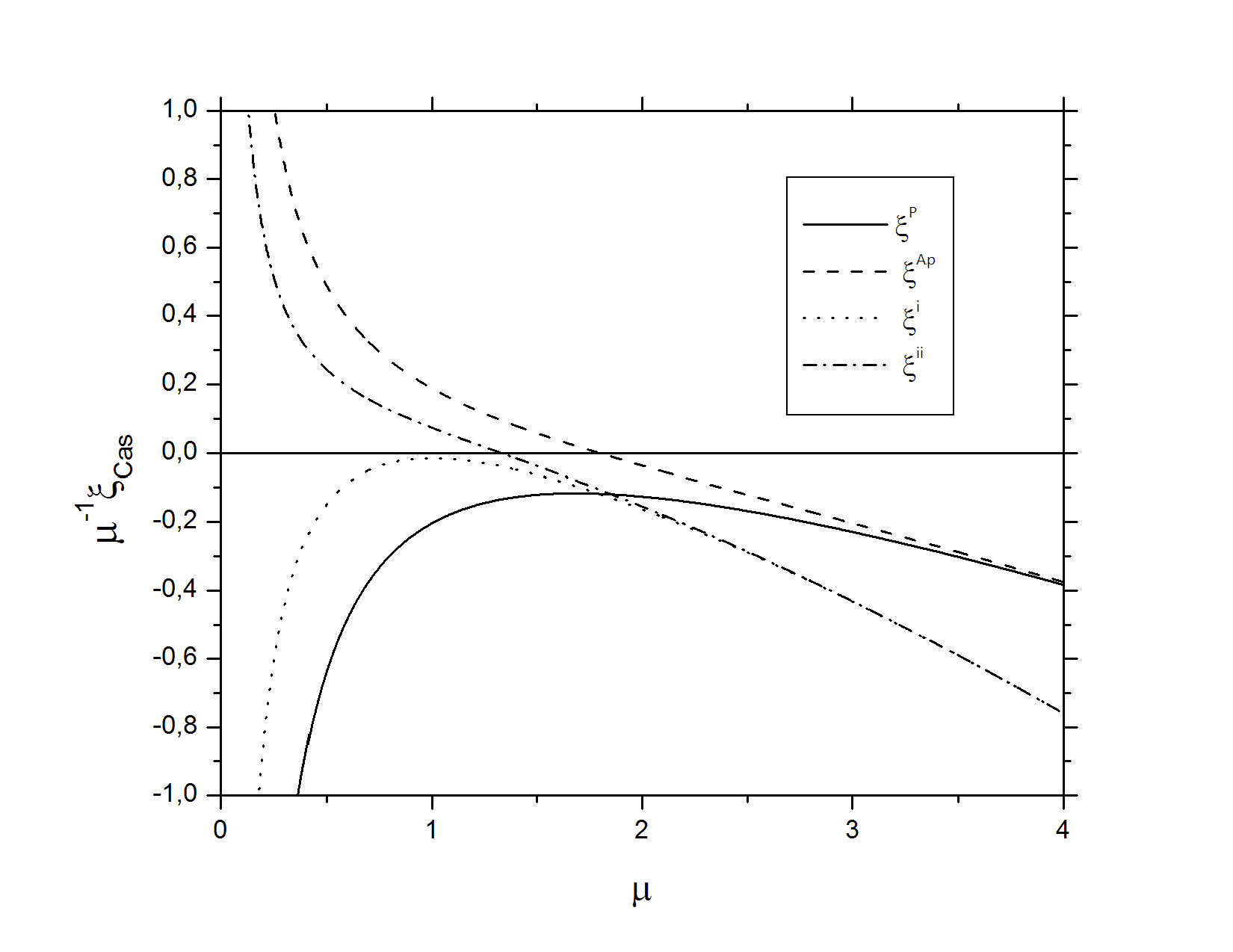}
\caption{Behaviour of $\xi$ for different BC's and $\tilde{\eta}=1$. We observe that $\xi^P$ and $\xi^i$ never reach the zero point energy.
We have an asymptotic behaviour coinciding $\xi^{P}$ with $\xi^{AP}$ and  $\xi^{i}$ with  $\xi^{ii}$.}
\label{fig-tc1}
\end{center}
\end{minipage}
\hfill
\begin{minipage}[t]{.45\textwidth}
\begin{center}
\includegraphics[width=100mm,scale=0.5]{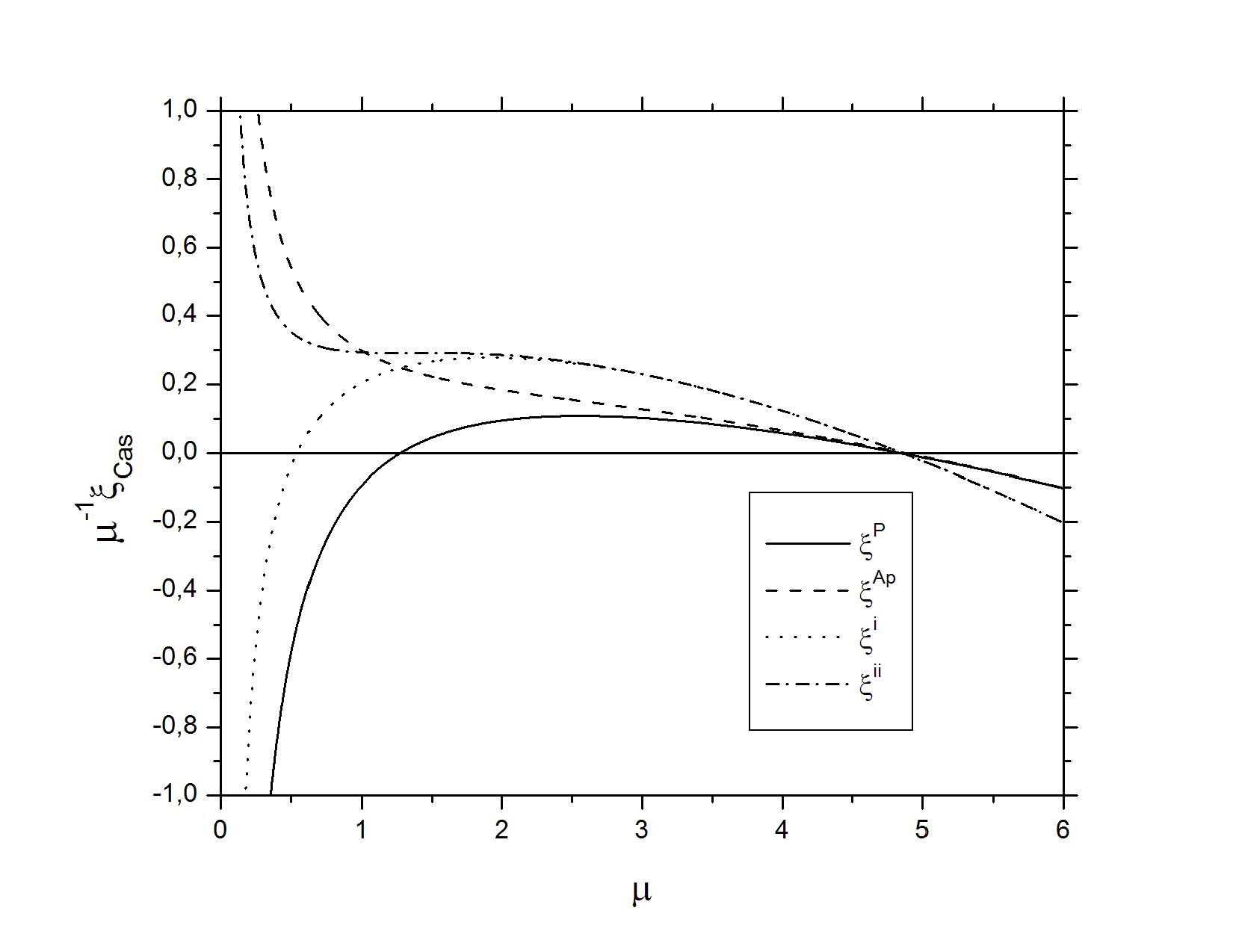}
\caption{Behaviour of $\xi$ for different BC's and $\tilde{\eta}=4$.
We obtain that that $\xi^P$ and $\xi^i$ crosses the zero point energy
for a certain region of the parameter $\mu$.}
\label{fig-tc2}
\end{center}
\end{minipage}
\hfill
\end{figure}

\noindent\textbf{Periodic BC}

\bigskip

Now we have the BC's 

\begin{eqnarray}
\label{cond.p}
f(x+L)& = &f(x),\nonumber\\
g(x+L)& = &g(x).
\end{eqnarray}

\noindent Proceeding as before, we obtain

\begin{equation}
\label{auto.p}
\lambda_{n}^{2}=m^{2}+\frac{4\pi^{2} n^{2}}{L^{2}}
\rightarrow \phi=0,\;r=1.
\end{equation}

\noindent The Casimir energy is given by

\begin{equation}
\label{energia.p}
{\xi}_{Cas}^{P}=-\frac{\mu^{2}}{8\pi}+\frac{\mu^{2}}{4\pi}\ln\left(\frac{2\tilde{\eta}}{\mu}\right)-\frac{\mu}{\pi}\sum_{n=1}^{\infty}\frac{1}{n}K_{1}(n\mu),
\end{equation}

\noindent and the Casimir force

\begin{equation}
\label{fuerzap}
\mathcal{F}^{P}_{Cas}=
\frac{1}{8\pi}
-\frac{1}{4\pi}\ln\left(\frac{2\tilde{\eta}}{\mu}\right)
-\frac{1}{\pi}\sum_{n=1}^{\infty}K_{0}(\mu n)
-\frac{1}{\mu \pi}\sum_{n=1}^{\infty}K_{1}(\mu n).
\end{equation}


\noindent\textbf{Zero current BC}

\bigskip

\bigskip

The confining condition is imposing the zero current condition at the borders

\begin{equation}
\label{cond.mit}
\left. i n^{\mu} \bar{\Psi}\gamma_{\mu}\Psi= 0\right |_{x=0}, \;\; \left. n^{\mu} \bar{\Psi}\gamma_{\mu}\Psi= 0\right |_{x=L}.
\end{equation}

\noindent And the eigenvalues are

\begin{eqnarray}
\label{auto.zc}
\lambda_{n}^{i,2}& = &m^{2}+\frac{n^{2}\pi^{2}}{L^{2}}
\rightarrow r=2 \;{\rm and }\;\phi=0,
\nonumber\\
\lambda_{n}^{ii,2}& = &m^{2}+\frac{(2n+1)^{2}\pi^{2}}{(2L)^{2}}
\rightarrow r=2 \;{\rm and }\;\phi=\pi.
\end{eqnarray}

According to ec.(\ref{ecas}) with $r=2$,  the Casimir energy and the Casimir force for this eigenvalues are given by

\begin{eqnarray}
\label{energia.zc}
{\xi}_{Cas}^{i}& = &-\frac{\mu^{2}}{4\pi}+\frac{\mu^{2}}{2\pi}\ln\left(\frac{2\tilde{\eta}}{\mu}\right)-\frac{\mu}{\pi}\sum_{n=1}^{\infty}\frac{K_{1}(2\mu n)}{n},\nonumber\\
{\xi}_{Cas}^{ii}& = &-\frac{\mu^{2}}{4\pi}+\frac{\mu^{2}}{2\pi}\ln\left(\frac{2\tilde{\eta}}{\mu}\right)-\frac{\mu}{\pi}\sum_{n=1}^{\infty}(-1)^{n}\frac{K_{1}(2\mu n)}{n}.
\end{eqnarray}

\begin{eqnarray}
\mathcal{F}^{i}_{Cas}& = &\frac{1}{4\pi}-\frac{1}{2\pi}\ln\left(\frac{2\tilde{\eta}}{\mu}\right)-\frac{2}{\pi}\sum_{n=1}^{\infty}K_{0}(2\mu n)-\frac{1}{\mu \pi}\sum_{n=1}^{\infty}\frac{K_{1}(2\mu n)}{n}, \nonumber\\
\mathcal{F}^{ii}_{Cas}& = &\frac{1}{4\pi}-\frac{1}{2\pi}\ln\left(\frac{2\tilde{\eta}}{\mu}\right)-\frac{2}{\pi}\sum_{n=1}^{\infty}(-1)^{n}K_{0}(2\mu n)-\frac{1}{\mu \pi}\sum_{n=1}^{\infty}(-1)^{n}\frac{K_{1}(2\mu n)}{n}.
\end{eqnarray}

\begin{figure}[h]
\hfill
\begin{minipage}[t]{.45\textwidth}
\begin{center}
    \includegraphics[width=100mm,scale=0.5]{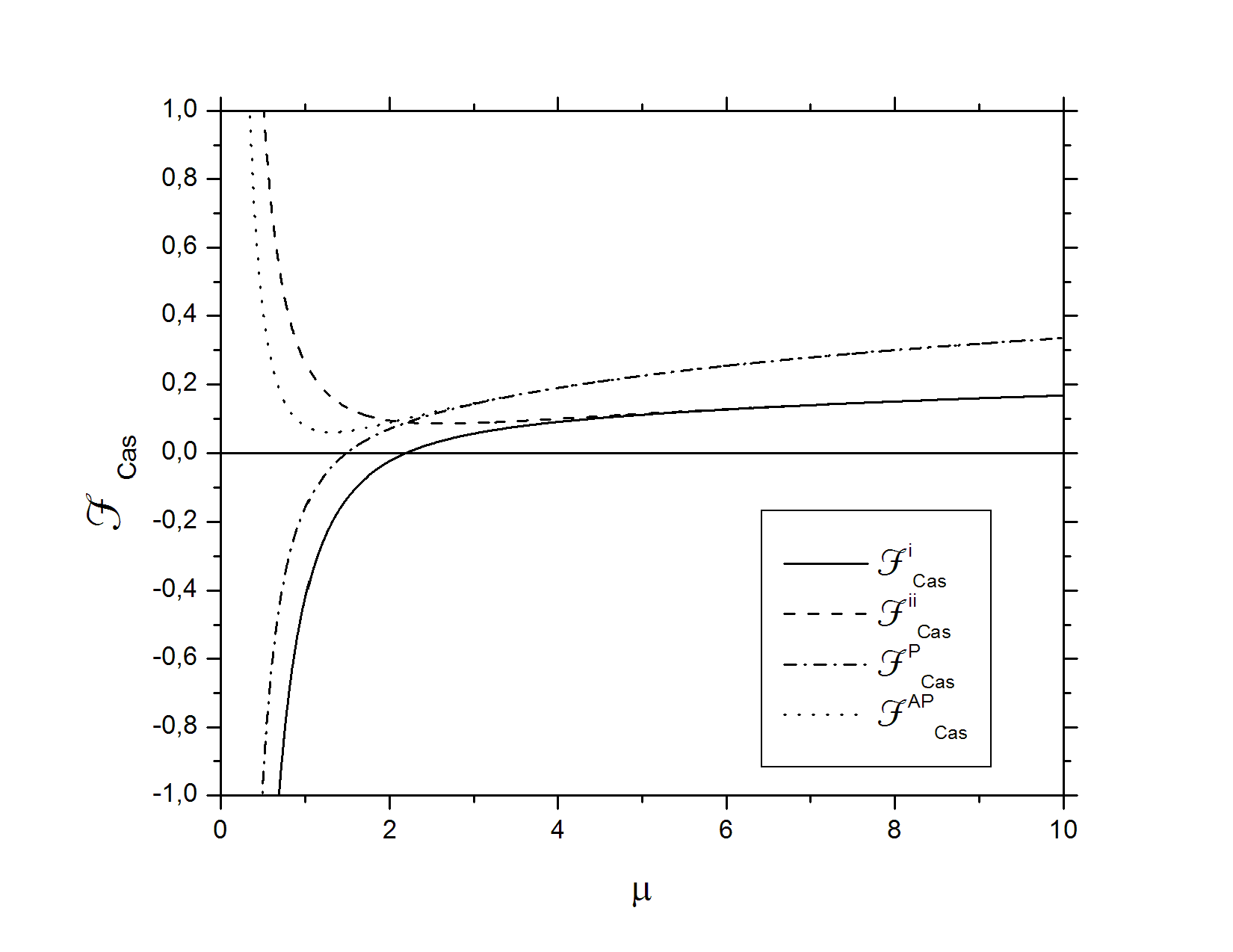}
\caption{Behaviour of ${\cal F}$ for different BC's and $\eta L=\tilde{\eta}=1$.
We can see that ${\cal F}^{AP}$ and ${\cal F}^{ii}$ are always positive.
It is also seen an asymptotic behaviour coinciding ${\cal F}^{P}$ with ${\cal F}^{AP}$ and  ${\cal F}^{i}$ with  ${\cal F}^{ii}$.}
\label{fig-f1}
\end{center}
\end{minipage}
\hfill
\begin{minipage}[t]{.45\textwidth}
\begin{center}
    \includegraphics[width=100mm,scale=0.5]{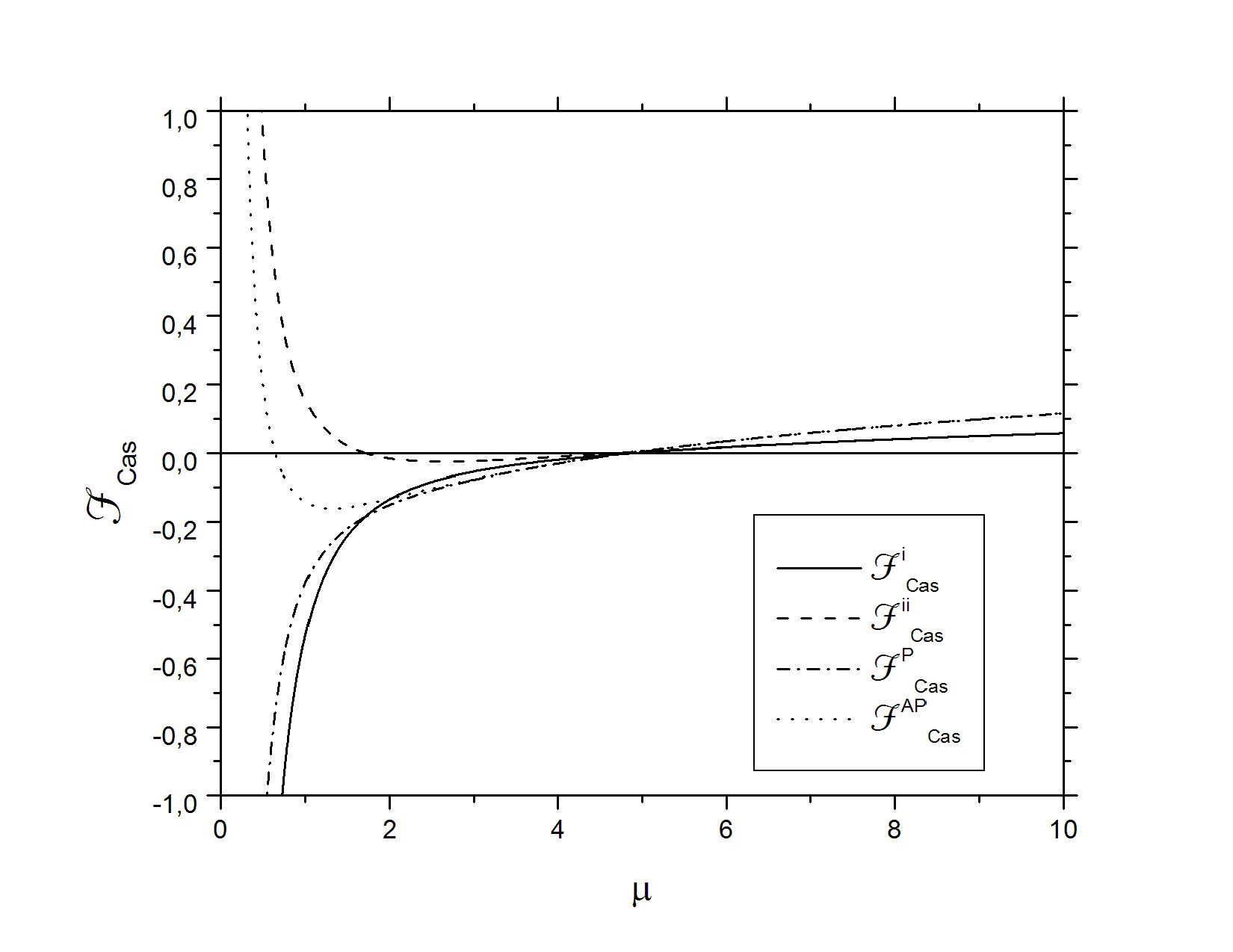}
\caption{Behaviour of ${\cal F}$ for different BC's and $\tilde{\eta}=4$. There it happens that ${\cal F}^{AP}$ and ${\cal F}^{ii}$ acquire a negative value in some limited region of $\mu$.}
\label{fig-f2}
\end{center}
\end{minipage}
\hfill
\end{figure}

\begin{figure}[h]
\hfill
\begin{minipage}[t]{.45\textwidth}
\begin{center}
    \includegraphics[width=100mm,scale=0.5]{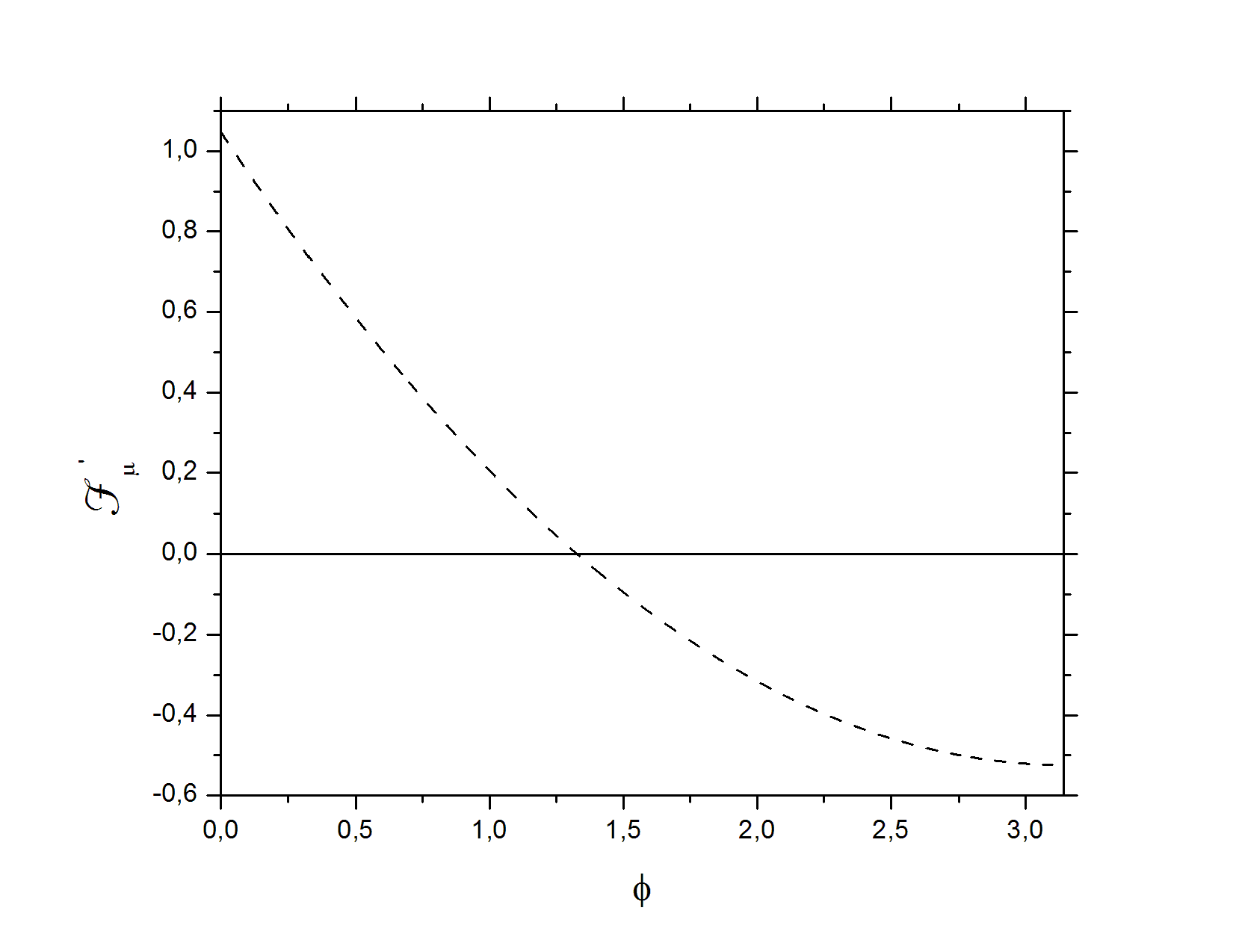}
\caption{Behaviour of the numerator in (\ref{df}). which indicates 
  the slope of the force when $\mu \approx 0$}
\label{dfdmu}
\end{center}
\end{minipage}
\hfill
\begin{minipage}[t]{.45\textwidth}
\begin{center}
    \includegraphics[width=100mm,scale=0.5]{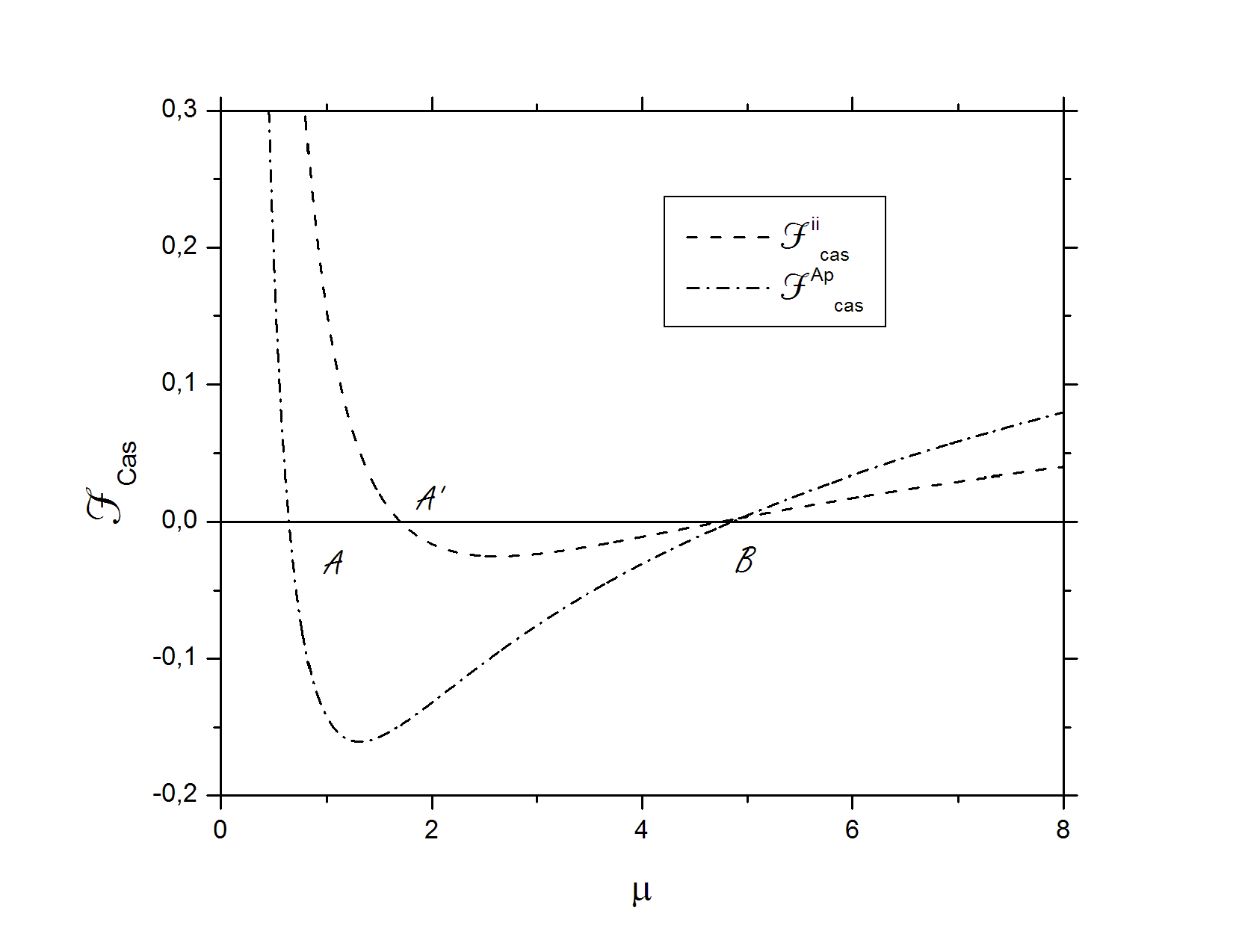}
\caption{Behaviour of ${\cal F}^{AP}$ and ${\cal F}^{ii}$ for $\tilde{\eta}=4$. There it happens that ${\cal F}^{AP}$ and ${\cal F}^{ii}$ acquire a negative value in region ${\cal A}({\cal A'})$ and becomes zero in the point ${\cal B}$.}
\label{puntosAB}
\end{center}
\end{minipage}
\hfill
\end{figure}


\noindent\textbf{Limiting values}

\bigskip

As can be seen from figures (\ref{fig-tc1}) and (\ref{fig-tc2}), the behaviour for
small $\mu$ depends on the BC's. In fact, the parameter 
$\phi$ determines the sign of the force as $\mu$ goes to zero.
We are interested in the sign of the force for $\mu \sim 0$, where 
the force clearly goes to $\pm \infty$. Keeping the leading terms 
for $\mu \approx 0$:

\begin{equation}
{\cal F} \approx -\frac{1}{\mu^2 \pi}\sum_{n=1}^{\infty}
\frac{\cos(\phi n)}{n^2}+{\rm constants}.
\end{equation}

\noindent It is more clear to take the derivative to leading order

\begin{equation}
\frac{d{\cal F}}{d\mu}_{\mu \rightarrow 0}\approx
\frac{1}{\mu^3\pi}
\left[
{\rm Li}_2({\rm e}^{i\phi})+{\rm Li}_2({\rm e}^{-i\phi})
\right],
\label{df}
\end{equation}

\noindent where ${\rm Li}_n(x)$ are Polylogarithm  functions (see, for example \cite{lewin}). Since the positive derivative means a negative force and vice versa.
The regime changes for the non physical value of $\phi =\phi^* \approx 1.328$, as it is shown in the figure (\ref{dfdmu}), notice that $\phi^*$ does not depend on $\tilde{\eta}$.

Another curious feature happen with ${\cal F}^{AP}$ and
${\cal F}^{ii}$. When $\eta$ goes beyond a given value
$\tilde{\eta}^*=\tilde{\eta}^*(\phi)$,
the force becomes negative, having an equilibrium points ${\cal A}({\cal A'})$
and a metastable point ${\cal B}$, as it is clear from figure (\ref{puntosAB}).


\section{Conclusions and discussion}

The first part of this letter was aware of the ultraviolet behaviour of he GN model for different BC's, in the framework of mean field theory assuming homogeneous solution and using zeta function regularization. We found that the beta function is independent of the type of boundary condition used, and that there appears a mass scale of arbitrary value.
The generated dynamical mass should depend on the BC's, if we have no prescription on the arbitrary mass scale.

Later, assuming, an homogeneous solution, we studied the Casimir energy and forces for different BC's, if we concentrate on the behaviour of $\xi/\mu$ from figures (\ref{fig-tc1}) and (\ref{fig-tc2}), we notice the following features:

\begin{enumerate}[a.]
\item Anti periodic: $\xi/\mu>0$ for $\mu< \mu^{\star}$ and $\xi/\mu<0$ for $\mu> \mu^{\star}$, for any positive value of $\tilde{\eta}$.
\item Periodic:  $\xi/\mu$ has a maximum value for a certain value of 
$\mu=\mu^{\dagger}$, having limiting values of $\xi/\mu \rightarrow \pm\infty$, for 
$\mu \rightarrow 0$. The sign of the maximum value of $\xi/\mu$, depends on the parameter $\tilde{\eta}$.
\item Confining i: The same qualitative behaviour of the periodic case.
\item Confining ii: $\xi/\mu>0$ for $\mu< \tilde{\mu}^{\star}$ and $\xi/\mu<0$ for $\mu> \tilde{\mu}^{\star}$, for any positive value of $\tilde{\eta}$, in a similar fashion as the anti periodic case.
\item We found that there is a common singular value of $\mu$ for  $\tilde{\eta} \geq 4$, where $\xi/\mu$ becomes zero.
\end{enumerate}

\bigskip

\noindent For the Casimir forces, from figures (\ref{fig-f1}) and (\ref{fig-f2}), we conclude that

\begin{enumerate}[a.]
\item Anti periodic BC: ${\cal F}^{AP}\rightarrow \infty$ for
$\mu\rightarrow 0$ and $\mu \rightarrow \infty$, for any value of $\tilde{\eta}$. It also happen that for $\tilde{\eta} \geq 4$, 
${\cal F}^{AP}$ can be negative in a finite range of $\mu$.
\item Periodic: ${\cal F}^{P} \rightarrow -\infty$ for $\mu\rightarrow 0$ and  ${\cal F}^{P} \rightarrow \infty$ for $\mu\rightarrow \infty$.
\item Confining i: It has the same qualitative behaviour as the periodic case.
\item Confining ii: It has the same qualitative behaviour as the anti periodic case.
\end{enumerate}

\noindent It is shown in figure (\ref{fig-f2}) that for $\tilde{\eta}\geq 4$, there is a common point $\mu$ where the Casimir force becomes zero for any boundary condition.

From the above considerations, we conclude that for BC's periodic and confining i, there are two regimes of forces, being negative for ``small'' $\mu$, representing an universe that has a shrinking tendency. 
On the other hand, when $\mu$ is ``big", our universe is an expanding one. 

For the anti periodic and confining ii, there is a more complex situation, since its behaviour depends on the value of $\tilde{\eta}$.
For $\tilde{\eta} \leq \tilde{\eta}^*$, the force is always positive, hence there is an expanding universe. For $\tilde{\eta} \geq 4$, there is mixed case as it is shown in figure (\ref{puntosAB}), there are  
the points ${\cal A},{\cal A'}$ and the universal point $B$. Between ${\cal A}$(${\cal A'}$) and $B$, the force becomes negative. It is also clear that $B$ is an unstable point and the points ${\cal A}$, ${\cal A'}$ are attracting points.
 
This study suggest that the natural further step is to consider a general relativity study where the spatial dynamics are affected by the quantum fluctuations of the Casimir energy and confirm if the BC's determine the existence of shrinking  or expanding low dimensional universes.


\appendix

\section{Epstein zeta function}
\label{apA}

We use an extended version of the Epstein zeta function is \cite{Kirsten:2010zp}

\begin{equation}
\label{eps}
\zeta_{E}(s;a,b)=\sum_{n=-\infty}^{\infty} \left(a^{2}+(n+b)^{2}\right)^{-s}.
\end{equation}

\noindent We can express the summation term, using the properties of
gamma function 

\begin{eqnarray}
\label{eps.gamma}
\sum_{n=-\infty}^{\infty} \left(a^{2}+(n+b)^{2}\right)^{-s}& = &\frac{1}{\Gamma (s)}\int_{0}^{\infty}t^{s-1}\sum_{n=-\infty}^{\infty}e^{-t((n+b)^{2}+a^{2})}dt,\nonumber\\
& = &\frac{1}{\Gamma (s)}\int_{0}^{\infty}t^{s-1}e^{-ta^{2}}\sum_{n=-\infty}^{\infty}e^{-t(n+b)^{2}}dt,
\end{eqnarray}

\noindent expression which is valid for $s\geq 1$. By means of the Jacobi inversion formulae \cite{Kirsten:2010zp}, we have

\begin{eqnarray}
\label{poison}
\sum_{n=-\infty}^{\infty}e^{-t(n+b)^{2}} &=& \sqrt{\frac{\pi}{t}}\sum_{n=-\infty}^{\infty}e^{-\frac{\pi^{2}n^{2}}{t}
-2\pi i b n}, \nonumber  \\
&=&\sqrt{\frac{\pi}{t}}+4\sqrt{\frac{\pi}{t}}
\sum_{n=1}^{\infty} e^{-\frac{\pi^{2}n^{2}}{t}}\cos  \left(2\pi b n\right)
\end{eqnarray}

\noindent So, we have

\begin{eqnarray}
\label{eps.expand}
\sum_{n=-\infty}^{\infty} \left(a^{2}+(n+b)^{2}\right)^{-s} & = &\left.\frac{\sqrt{\pi}}{\Gamma(s)}\right[
\int_{0}^{\infty}t^{s-3/2}e^{-ta^{2}}dt \nonumber\\
&&+\left.  \quad 2 \sum_{n=1}^{\infty} \cos \left(2\pi b n\right)
\int_{0}^{\infty}t^{s-3/2}e^{-ta^{2}-\pi^{2}n^{2}/t}dt\right].
\end{eqnarray}

\noindent The above integrals are easily recognized \cite{gradshteyn} and have the form

\begin{eqnarray}
\int_{0}^{\infty}x^{\alpha-1}e^{-\gamma x}dx &=& \gamma^{-\alpha}\Gamma(\alpha),
\\
\int_{0}^{\infty} x^{\alpha-1}e^{-\beta/x-\gamma x}dx &=& 2\left(\frac{\beta}{\gamma}\right)^{\alpha/2}K_{\alpha}(2\sqrt{\beta\gamma}).
\end{eqnarray}

\noindent Leading us to the general expression

\begin{equation}
\label{c.a.eps}
\sum_{n=-\infty}^{\infty} \left(a^{2}+(n+b)^{2}\right)^{-s}  =
\frac{\sqrt{\pi}}{\Gamma(s)}\left[a^{-2(s-1/2)}\Gamma(s-1/2)
+4\sum_{n=1}^{\infty}\cos(2\pi b n) 
\left(\frac{\pi n}{a} \right)^{(s-1/2)}
K_{s-1/2}(2\pi a n) \right].
\end{equation}

\noindent We are interested in the case $s=-1/2$, but there is a singularity in such point, so we isolate it by computing for the
value $s=-1/2+\epsilon$, giving the expression

\begin{equation}
\label{}
\sum_{n=-\infty}^{\infty} \left(a^{2}+(n+b)^{2}\right)^{-s}  =
\sqrt{\pi} \left[
a^2a^{-2\epsilon}\frac{\Gamma(-1+\epsilon)}{\Gamma(-1/2+\epsilon)}+\frac{4 a}{\pi \Gamma(-1/2+\epsilon)}
\sum_{n=1}^{\infty}\frac{\cos(2\pi b n)}{n} K_{1}(2\pi a n)
 \right].
\end{equation}

\noindent In order to isolate the $\epsilon$ term, we use

\[
\sqrt\pi a^2a^{-2\epsilon}\frac{\Gamma(-1+\epsilon)}{\Gamma(-1/2+\epsilon)}=\sqrt\pi a^2
e^{-2\epsilon\ln(a)}\frac{\Gamma(-1+\epsilon)}{\Gamma(-1/2+\epsilon)} \approx
\frac{a^2}{2\epsilon}-\frac{a^2}{2}-a^2\ln\left(\frac{a}{2} \right)
\]

\noindent Finally, using the fact that $K_{n}(x)=K_{-n}(x)$ the Epstein function can be expressed in a term where
the singular point becomes isolated

\begin{equation}
\label{}
\zeta_{E}\left(\epsilon;a,b\right)=\frac{a^{2}}{2\epsilon}
-\frac{a^{2}}{2}-a^{2}\ln\left(\frac{a}{2}\right)-\frac{2 a}{\pi}\sum_{n=1}^{\infty} \frac{\cos(2\pi b n)}{n} K_{1}(2\pi a n).
\end{equation}

\noindent The limit $a\rightarrow 0$ of the finite part

\begin{eqnarray}
\lim_{a\rightarrow 0} FP\zeta_{E}\left(\epsilon;a,b\right)
&=& -\frac{1}{2\pi^ 2}\left[
\rm{Li}_2(e^{i2\pi b})+\rm{Li}_2(e^{-i2\pi b})
\right] \nonumber \\
&=&  -\frac{1}{2\pi^ 2}\left[
\frac{\pi^2}{6}-\frac{2\pi^2 b}{2}+\frac{4\pi^2b^2}{4}
\right] 
= 
-\frac{1}{12}+\frac{b}{2}-\frac{b^2}{2}.
\end{eqnarray}


\section{Computation of energy density for general BC's}
\label{apB}

 As we see from section II, the density of energy 

\[
\frac{\mathcal{E}}{N}=-\frac{2}{r L}\sum_{n} \left(m^{2}+k_{n}^{2}\right)^{1/2} +\frac{m^{2}}{2G},
\]

\noindent is given by the BC's imposed over $k_{n}$ 

\[
k_{n}^{2}=\frac{(2\pi n +\phi)^{2}}{r^{2}L^{2}}.
\]

\noindent We can use the zeta function regularization in order to obtain an expression for the energy density

\begin{eqnarray}
\label{}
\frac{\mathcal{E}}{N}& = &-\frac{2}{r L}\sum_{n=-\infty}^{\infty} \left(m^{2}+\frac{(2\pi n +\phi)^{2}}{r^{2}L^{2}}\right)^{-s} +\frac{m^{2}}{2G},\nonumber\\
& = &-\frac{2}{rL}\left(\frac{2\pi}{rL}\right)^{-2s}\sum_{n=-\infty}^{\infty}\left(\frac{m^{2}L^{2}r^{2}}{4\pi^{2}}+\left(n+\frac{\phi}{2\pi}\right)^{2}\right)^{-s}+\frac{m^{2}}{2G}.
\end{eqnarray}

\noindent  Introducing a parameter of mass $\eta$ and $L^{2}$  in both sides of the equation

\begin{equation}
\label{}
\frac{L^{2}\mathcal{E}}{N}=-\frac{2L}{r}\eta^{2s+1}\left(\frac{2\pi}{rL}\right)^{-2s}\sum_{n=-\infty}^{\infty}\left(\frac{\mu^{2}r^{2}}{4\pi^{2}}+\left(n+\frac{\phi}{2\pi}\right)^{2}\right)^{-s}+\frac{\mu^{2}}{2G}.
\end{equation}

\noindent Recognizing the sum as the Epstein zeta function (see appendix A) and with $s=\epsilon-1/2$, we have

\[
\frac{L^{2}\mathcal{E}}{N}=-\frac{4\pi}{r^{2}}\left(\frac{\eta L r}{2\pi}\right)^{\epsilon}\left[\frac{\mu^{2}r^{2}}{8\pi^{2}\epsilon}-\frac{\mu^{2}r^{2}}{8\pi^{2}}-\frac{\mu^{2}r^{2}}{4\pi^{2}}\ln\left(\frac{\mu r}{4\pi}\right)-\frac{\mu r}{\pi^{2}}\sum_{n=1}^{\infty}\frac{\cos \phi n}{n}K_{1}(\mu r n)\right]+\frac{\mu^{2}}{2G},
\]

\noindent for $\epsilon\to 0$ the energy density is given by

\begin{equation}
\label{den}
\frac{\mathcal{E}}{N}=-\frac{\mu^{2}}{2\pi\epsilon L^{2}}+\frac{\mu^{2}}{2\pi L^{2}}-\frac{\mu^{2}}{\pi L^{2}}\ln \frac{2\tilde{\eta}}{\mu}+\frac{4\mu}{\pi r L^{2}}\sum_{n=1}^{\infty}\frac{\cos \phi n}{n}K_{1}(\mu r n)+\frac{\mu^{2}}{2G L^{2}},
\end{equation}

\noindent where $\tilde{\eta}\equiv \eta L$.

\noindent Minimizing the energy density respect to $\mu$ 

\begin{equation}
\label{emin}
\frac{1}{N}\frac{\partial \mathcal{E}}{\partial \mu}=-\frac{\mu}{\pi\epsilon L^{2}}+\frac{2\mu}{\pi L^{2}}-\frac{2\mu}{\pi L^{2}}\ln\left(\frac{2\tilde{\eta}}{\mu}\right)-\frac{4\mu}{\pi L^{2}}\sum_{n=1}^{\infty}\cos (\phi n) K_{0}(\mu r n)+\frac{\mu}{G L^{2}},
\end{equation}

\noindent and considering that $(\partial \mathcal{E})/\partial \mu=0$, we can obtain an dimensionless expression 

\[
\frac{L^{2}}{N \mu}\frac{\partial \mathcal{E}}{\partial \mu}\equiv \mathcal{X}_{\mu}=-\frac{1}{\pi\epsilon}+\frac{2}{\pi}-\frac{2}{\pi}\ln\left(\frac{2\tilde{\eta}}{\mu}\right)-\frac{4}{\pi}\sum_{n=1}\cos(\phi n)K_{0}(\mu r n)+\frac{1}{G}=0,
\]

\noindent therefore the parameter $G$ is given by

\begin{equation}
\label{G}
G=\pi\left\{\frac{1}{\epsilon}-2+2\ln\left(\frac{2\tilde{\eta}}{\mu}\right)+\frac{4}{\pi}\sum_{n=1}^{\infty}\cos(\phi n)K_{0}(\mu r n)\right\}^{-1}.
\end{equation}


\section{No current through the boundary}
\label{apC}

It is imposed the zero current condition at the boundaries

\begin{equation}
\label{cond.mit}
\left. i n^{\mu} \bar{\Psi}\gamma_{\mu}\Psi= 0\right |_{x=0}, \;\; \left. n^{\mu} \bar{\Psi}\gamma_{\mu}\Psi= 0\right |_{x=L}.
\end{equation}

\noindent In terms of components, we have 

\[
\psi={\phi(x)  \choose \chi(x)} 
\rightarrow n^{\mu} \bar{\Psi}\gamma_{\mu}\Psi=\phi(x)\chi(x)^*-\phi(x)^*\chi(x).
\]

\noindent if $
\phi(x)= \left|\phi(x)\right| {\rm e}^{\alpha},\;\chi(x)= \left|\chi(x)\right| {\rm e}^{\beta}$, then

\[
i n^{\mu} \bar{\Psi}\gamma_{\mu}\Psi=2 i \left|\phi(x)\right|
 \left|\chi(x)\right| \sin(\alpha-\beta)
\]

\noindent Since $\alpha$ and $\beta$ are constants, we must impose that at the borders one of the fields must be zero, we can consider the following cases:

\begin{itemize}

\item[i)]  $\chi(0)=0, \;\chi(L)=0$ or 
 $\phi(0)=0, \;\phi(L)=0$,

\item[ii)] $\phi(0)=0, \;\chi(L)=0$ or 
 $\chi(0)=0, \;\phi(L)=0$.
 
\end{itemize}

\noindent The conditions are

\begin{itemize}

\item[i)] $\sin\left( \sqrt{\lambda^2-m^2}L\right)=0
\rightarrow \lambda_n^2=m^2+\left(\frac{n\pi}{L}\right)^2
$.

\item[ii)] $\sqrt{\frac{\lambda^2}{m^2}-1}\cos\left(\sqrt{\frac{\lambda^2}{m^2}-1}\mu\right)-
\sin\left(\sqrt{\frac{\lambda^2}{m^2}-1}\mu\right)=0
$, a transcendental equation that for $n\rightarrow \infty$
behaves as $\lambda_n^2=m^2+\left(\frac{(2n+1)\pi}{2L}\right)^2$.
 
\end{itemize}


\bigskip

{\bf Acknowledgments}\\[.2cm]
F.E. and J.C.R. aknowledge the support of  FONDECYT under grant No. 1150471 and J.C.R. aknowledges support of FONDECYT under grants No. 1150847 and No. 1130056.



\begin{thebibliography}{10}

\bibitem{Gross:1974jv}
  D.~J.~Gross and A.~Neveu,
  Phys.\ Rev.\ D {\bf 10}, 3235 (1974).


\bibitem{Thies:2005vq}
  M.~Thies and K.~Urlichs,
  Phys.\ Rev.\ D {\bf 71}, 105008 (2005)
  [hep-th/0502210].

\bibitem{saxena}
A.~Saxena and A.~R.~Bischop,
  Phys.\ Rev.\ A {\bf 44}, R2251 (1991).

\bibitem{brazovskii}
S.~A.~Brazovskii and N.~N.~Kirova,
 JETP \ Lett.\ {\bf 33}, 4 (1981).
 

\bibitem{Thies:2005wv} 
  M.~Thies and K.~Urlichs,
  Phys.\ Rev.\ D {\bf 72}, 105008 (2005)
  doi:10.1103/PhysRevD.72.105008
  [hep-th/0505024].


\bibitem{Kirsten:2010zp} 
  K.~Kirsten,
  MSRI Publ.\  {\bf 57}, 101 (2010)
  [arXiv:1005.2389 [hep-th]].

 
\bibitem{Schon:2000he} 
  V.~Schon and M.~Thies,
  Phys.\ Rev.\ D {\bf 62}, 096002 (2000)
  doi:10.1103/PhysRevD.62.096002
  [hep-th/0003195].
 


\bibitem{gradshteyn} I. S. Gradshteyn and I. M. Ryzhik, \emph{Table of integrals, series and products},(7th
Ed.) Academic Press, New York, 1980

\bibitem{lewin} Lewin (1981), \emph{Polylogarithms and Associated Functions}. North-Holland Publishing Co., New York, 1981 

\end{thebibliography}
\end{document}